\documentstyle[multicol,aps,amssymb,epsfig]{revtex}

\begin{document}

\draft
\title{Fluctuation--dissipation relations in the activated regime
of simple strong--glass models}

\author{Arnaud Buhot and Juan P. Garrahan}

\address{Theoretical Physics, University of Oxford, 1 Keble Road,
Oxford, OX1 3NP, U.K.}

\date{November 2, 2001}

\maketitle

\begin{abstract}
We study the out--of--equilibrium fluctuation--dissipation (FD)
relations in the low temperature, finite time, physical aging regime
of two simple models with strong glass behaviour, the
Fredrickson--Andersen model and the square--plaquette interaction
model. We explicitly show the existence of unique, waiting--time
independent dynamical FD relations. While in the Fredrickson--Andersen
model the FD theorem is obeyed at all times, the plaquette model
displays piecewise linear FD relations, similar to what is found in
disordered mean--field models and in simulations of supercooled
liquids, and despite the fact that its static properties are
trivial. We discuss the wider implications of these results.
\end{abstract}

\vspace{-0.1cm}


\vspace{-0.2cm}

\begin{multicols}{2}
\narrowtext

The common feature to all glassy systems, like supercooled liquids,
spin glasses, and, to a certain degree, even soft materials like
gently driven powders, is an extremely slow relaxational dynamics at
low temperatures or high densities (for reviews see
\cite{Angell,Bouchaud,Debe}).  Often, glassy systems display a
dynamical behaviour known as aging, which corresponds to the
asymptotic regime in which one--time quantities, like energy or
magnetization, are stationary, but two--time quantities, like response
and autocorrelation functions, still depend on the time elapsed since
the system was prepared, rather than just on time differences, as in
equilibrium. While in this situation response and autocorrelations do
not obey the fluctuation--dissipation theorem (FDT) \cite{Chandler},
exact results for mean--field models \cite{CuKu,Bouchaud} have
suggested that FD relations are generalized in a well defined way,
that the breakdown of FDT can be understood in terms of `effective'
temperatures \cite{CuKuPe}, and that there may be a close connection
between out--of--equilibrium FD relations and equilibrium properties
\cite{Franz}. Nontrivial asymptotic FD relations have also been found
in other (non--glassy) out--of--equilibrium situations, like
ferromagnetic domain growth \cite{Barrat} and systems at criticality
\cite{Godreche}, and FD plots similar to the ones of discontinuous
mean--field models have been observed in simulations of supercooled
liquids \cite{Parisi,Kob} and of frustrated \cite{Ricci} and
constrained \cite{BarratKurchan} lattice gases.

However, both in experiments and simulations, a relevant regime is
that of long but {\em finite} times where one--time quantities are not
stationary, but are slowly relaxing towards their equilibrium values,
a situation known as `physical' aging \cite{Struik}. A second issue is
that near the glass transition activated processes, which are
explicitly excluded in mean--field, play an essential role. With
respect to this, simulations of simple models with activated dynamics
\cite{Crisanti,Garrahan} have shown nonmonotonic response functions,
which, superficially, lead to meaningless FD relations (something
analogous occurs in models of vibrated granular matter
\cite{Nicodemi}).  And finally, there is evidence, at least for
molecular glasses, for the absence of any thermodynamic phase
transition underlying the dynamical arrest \cite{Santen}, so that if
nontrivial FD relations exist for these systems they cannot be
interpreted in terms of the structure of an equilibrium glass phase.

The purpose of this Letter is to address the problem of whether well
defined out--of--equilibrium FD relations can be obtained when all of
the above factors are taken into account. We do this by considering
the case of two simple non--disordered or frustrated systems with
trivial statical properties but dynamical behaviour characteristic of
a strong--glass \cite{Angell} (like Arrhenius relaxation time,
exponential relaxation functions, etc.): the 1D and 2D
Fredrickson--Andersen model \cite{Fredrickson,Schulz,Crisanti,Buhot}
and the 2D square--plaquette interaction model \cite{Plaquette}. We
show that for an appropriately defined class of observables
nontrivial, unique out--of--equilibrium FD relations exist in the
physical aging regime of these models. Using simple scaling
arguments we find that in the FA model FDT is obeyed at all
times. Remarkably, in the plaquette model the existence of two
relevant timescales for the relaxation leads to piecewise linear FD
relations, despite its trivial statics.

Let us consider first the case of the Fredrickson--Andersen (FA) model
\cite{Fredrickson}, which corresponds to Ising spins $\sigma_i = \pm
1$ $(i=1, \dots, N)$, with Hamiltonian $H = \sum_i n_i$, where $n_i
\equiv (1+\sigma_i)/2$, and subject to a single spin--flip dynamics
with the kinetic constraint that only spins which have at least one
nearest neighbour in the up state are allowed to flip. This dynamics
obeys detailed balance, and the equilibrium is trivial.  The energy
density is given by the concentration $c$ of up spins (or `defects'),
which in equilibrium becomes $c_{\rm eq}=1/(1+e^{1/T})$. At low
temperatures $c$ is very small, and since defects facilitate the
dynamics, the system slows down. Isolated defects are locally stable
and the system has to overcome energy barriers to evolve. There is a
single activation barrier to the diffusion of defects $\Delta E=1$,
which implies that relaxation times follow the Arrhenius law
$\tau_{\rm FA} \sim e^{{\rm const.}/T}$, characteristic of strong
glass behaviour \cite{Buhot}.  We wish to study the dynamical FD
relations for long but finite times after a quench from $T_0=\infty$
to a low temperature $T$. This is the physical aging regime of the
system in which one--time quantities change slowly with time as they
relax towards their equilibrium values. For low temperatures, the
concentration of defects $c(t)\equiv N^{-1} \sum_i \langle n_i(t)
\rangle$ develops a plateau at $c_p$, which becomes longer the lower
the temperature, and corresponds to the onset of activation. The
plateau is reached at $t \sim 1$ and this initial transient is $T$
independent. We are interested in times $t > 1$, that is, from the
plateau onwards, for which the relaxation proceeds through activated
processes.

The integrated response for an individual spin can be measured by the
now standard method \cite{Barrat} of applying a random field
perturbation, $\delta H(t) = -h(t) \sum_i \varepsilon_i n_i$, where
$\varepsilon_i=\pm1$ are iid random variables. For a constant field
applied from waiting time $w$ onwards, the integrated response at time
$t$ is given by $\chi(t,w) \equiv (Nh)^{-1} \sum_i
\overline{\varepsilon_i \langle n_i(t) \rangle_h}$, where $\langle
\cdot \rangle_h$ indicates dynamical average in the presence of the
perturbation, and $\overline{(\cdot)}$, average over the realization
of the random field.  In Fig.\ 1 (top left) we show $\chi(t,w)$ for
low temperatures and various waiting times for the 1D FA model
\cite{Simu}, the behaviour of the 2D version being similar. Notice
that the response function is non--monotonic \cite{Crisanti}. The peak
in the response decreases with increasing waiting time, and the
function eventually becomes monotonic when the system equilibrates. On
the other hand, the spin autocorrelation function $C(t,w) \equiv
N^{-1} \sum_i \langle n_i(t) n_i(w) \rangle$ is monotonically
decreasing, as expected.  The combination of non--monotonic response
and monotonic correlation was interpreted as leading to meaningless FD
relations for this model \cite{Crisanti}.  A more careful analysis
reveals something different.  Only defects contribute to the response
function which implies that $\chi(t,w)$ should be proportional to
$c(t)$. Moreover, to leading order in $e^{-1/T}$, a spin which can
respond is isolated and, thus, in {\it local} equilibrium. This leads
to the following scaling form for the response for $T < 1$ and $w >
1$,
\begin{equation}
\chi(t,w) \sim \frac{c(t)}{c_{\rm eq}} \chi_{\rm
eq}\left[\frac{(t-w)}{e^{1/T}} \right] ,
\label{resp1}
\end{equation}
where $\chi_{\rm eq}$ is the equilibrium response function, which
depends only on the time difference, and scales with the temperature
through the relaxation time of up spins. Fig.\ 1 (top right) 
presents the excellent collapse of all the data of Fig.\ 1 (top left)
under Eq.\ (\ref{resp1}). The above expression also reveals the nature
of the non--monotonicity in the out--of--equilibrium susceptibility:
it is the product of a decreasing function, $c$, and an increasing
one, $\chi_{\rm eq}$. Similar arguments give the scaling form of the
autocorrelation $C(t,w)$: of the up spins at the earlier time, $c(t)$
will remain and $c(w)-c(t)$ will disappear. Assuming that the
autocorrelation of the former is proportional to the equilibrium one,
and that they are uncorrelated to the latter, we obtain,
\begin{equation}
C(t,w) \sim \frac{c(t)}{c_{\rm eq}} C_{\rm eq}\left[\frac{(t-w)}{e^{1/T}}
\right] + c(t) [c(w)-c(t)],
\label{corr1}
\end{equation}
where $C_{\rm eq}$ is the equilibrium autocorrelation function. Fig.\
1 (bottom left) shows the excellent collapse of autocorrelation
functions under (\ref{corr1}).

\begin{figure}[t]
\begin{center}
\epsfig{file=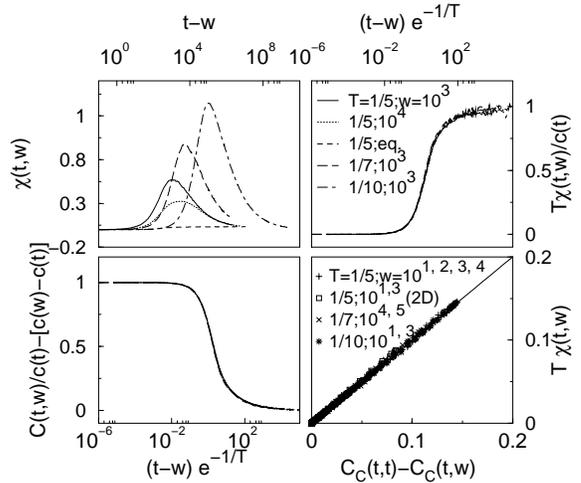, width=3.0in}
\caption
{Top left: integrated response $\chi(t,w)$ vs.\ time difference $t-w$
in the 1D FA model. Top right and bottom left: rescaled response Eq.\
(\ref{resp1}) and rescaled autocorrelation function Eq.\ (\ref{corr1})
as a function of scaled time difference $(t-w) e^{-1/T}$. Bottom
right: out--of--equilibrium FD relations in the 1D and 2D FA models
obtained from parametric plots of response vs.\ difference of
connected autocorrelation.  }
\end{center}
\end{figure}

\vspace{-0.3cm}

Having expressed the out--of--equilibrium response and correlation
functions in terms of the equilibrium ones, we can now find the FD
relations. Since we know that the equilibrium functions satisfy the FD
theorem (FDT) \cite{Chandler}, $T \chi_{\rm eq}(\tau) = C_{\rm eq}(0)
- C_{\rm eq}(\tau)$, from Eqs.(\ref{resp1}) and (\ref{corr1}) we
obtain,
\begin{equation}
T \chi(t,w) \sim C_c(t,t) - C_c(t,w) ,
\label{fdt1}
\end{equation}
where $C_c(t,t') \equiv C(t,t') - c(t) c(t')$ is the {\em connected}
autocorrelation function \cite{Connected}. This means that for all low
$T$ and $w > 1$ there is a unique dynamical FD relation which
corresponds to the out--of--equilibrium response and autocorrelation
obeying FDT. We illustrate this for various temperatures and waiting
times in Fig.\ 1 (bottom right) for the 1D and 2D models. Notice that
this result does not mean that the system is in equilibrium: one--time
quantities, like the defect concentration $c(t)$, are orders of
magnitude away from their equilibrium values; and, moreover, from the
plateau onwards, the system develops a strong nearest--neighbour
repulsion, $\langle \delta n_i(t) \delta n_{i \pm 1}(t) \rangle \sim -
c(t) [c(t) - c_{\rm eq}]$, which is a manifestation of the fact that
the dynamics explores mostly local minima, and eventually
vanishes in equilibrium. In fact, the response to a uniform field and
the autocorrelation of the magnetization are not related by any
sensible FD relation.

Let us turn now to the more interesting case of the two--dimensional
square--plaquette model, which consists of a system of Ising spins,
$\sigma_i = \pm 1$ $(i=1, \dots, N)$, in a square lattice, with
ferromagnetic interactions between quartets of neighbouring spins in
the vertices of the plaquettes of the lattice, $H = - \sum_{ijkl \in
\square} \sigma_i \sigma_j \sigma_k \sigma_l$.  This 
model is a special case of the eight vertex model \cite{Baxter} with
trivial thermodynamical properties, but whose single spin--flip

\begin{figure}[t]
\begin{center}
\epsfig{file=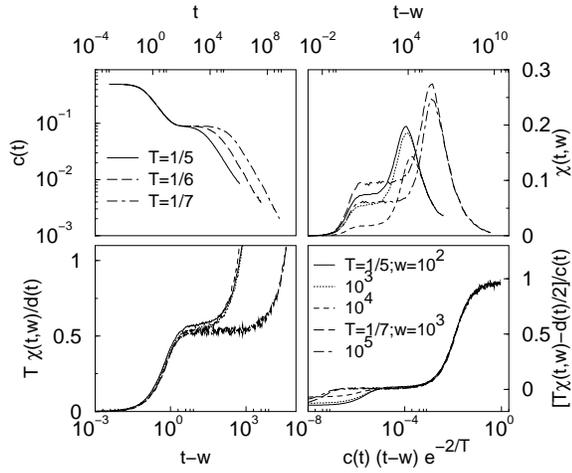, width=3.0in}
\caption
{Top left: concentration of defects $c(t)$ as a function of time after
the quench to $T$ in the 2D square--plaquette model.  Top right:
integrated response $\chi(t,w)$ as a function of time difference
$t-w$. Bottom left: rescaled form of response for early times,
Eq.(\ref{resp2a}). Bottom right: rescaled response for late times,
Eq.(\ref{resp2b}), vs.\ rescaled time $c(t) (t-w) e^{-2/T}$.}
\end{center}
\end{figure}

\vspace{-0.4cm}

\noindent
dynamics is
glassy \cite{Plaquette,Mila}. The model has a dual representation in
terms of noninteracting plaquette variables, $b_i \equiv \sigma_i
\sigma_j \sigma_k \sigma_l$. Since reversing any spin corresponds to
inverting the four plaquettes to which it belongs, the spin--flip
dynamics becomes a constrained dynamics in the plaquette
representation \cite{Plaquette}.

Isolated excitations or defects ($b_i=-1$) in the plaquette model are
stable, and have to overcome an energy barrier to move. In contrast to
the FA model, however, pairs of neighbouring defects can move at no
energy cost: a horizontal (vertical) pair can diffuse freely in a
vertical (horizontal) direction, while a diagonal pair cannot diffuse
but is free to oscillate. These excitations play an important role. An
isolated defect moves by creating a diffusing pair, so that $\Delta
E=2$.  The single activation barrier means that this model also
behaves as a strong glass.  Fig.\ 2 (top left) displays the decay of
the concentration of defects $c(t)\equiv N^{-1} \sum_i \langle n_i(t)
\rangle$, where $n_i \equiv (1-b_i)/2$, as a function of time after a
quench from a random state to low temperatures \cite{Simu}. As a
consequence of the presence of the diffusing pairs the plateau is
reached more slowly than in the FA model. The activated regime
corresponds to times $t > 10^2$ and $T < 0.25$ in this case.  At the
plateau the defect concentration is $c_P=0.089$, that of moving pairs
is $O(e^{-1/T} c_P)$, and that of defects which belong to oscillating
pairs is finite, $d_P=0.026$.

In analogy with the FA model, we now consider the linear response of
the excitations $n_i$, instead of that of the spins $\sigma_i$ which
in this class of models do not obey any systematic
out--of--equilibrium FD relations \cite{Garrahan}.  In Fig.\ 2 (top
right) we show the integrated response function of individual
excitations, $\chi(t,w)$, for $w>10^2$ and low temperatures.  The
response is again non--monotonic, but in this case it also presents an
intermediate saturation at early time lag. This is a consequence of
the existence of two well separated relevant timescales. The early
behaviour is due to the fast response of the oscillating pairs, which
exist in finite number in the activated regime. This means that we
expect, to leading order in $e^{-1/T}$, the scaling form of the
response function for low $T$, $w>10^2$, and $t-w \sim O(1)$ to be,
\begin{equation}
T \chi(t,w) \sim d(t) \, f(t-w) ,
\label{resp2a}
\end{equation}
where $d(t)$ is the number of excitations which belong to an
oscillating pair as a function of time, and $f(\tau)$ is an
increasing, temperature independent, function, which should saturate
at a half, $f(\infty)=1/2$, given that each oscillator can be in two
states. The collapse under (\ref{resp2a}) of the response curves for
early time lags is given in Fig.\ 2 (bottom left).  After the
saturation of the oscillators, the response of the isolated defects
takes over, so we expect the response function to be proportional to
$c(t)$, and the time lag to scale with the relaxation time, so that
for $t-w \gg 1$,
\begin{equation}
T \chi(t,w) \sim \frac{1}{2} \, d(t) + c(t) \, g\left[ c(t)
\frac{(t-w)}{e^{2/T}}\right] ,
\label{resp2b}
\end{equation}
where $g(\tau)$ is a temperature independent function with
$g(\infty)=1$, and the relaxation time $\tau_{rel} \sim 
e^{2/T}/c(t)$ for the diffusion of isolated defects scales 
with time through $c(t)$ to account for the necessary number 
of diffusing pairs created to reach another isolated defect.
This rather unexpected behaviour comes from the restricted 1D 
motion of the diffusing pairs and leads to a relaxation time 
$\tau_{rel} \sim e^{3/T}$ in equilibrium.  
In Fig.\ 2 (bottom right) we show that (\ref{resp2b}) accurately 
scales all the response curves for long time lags. Similar scaling 
relations can be obtained for the autocorrelation function.

The existence of two relevant timescales for the out--of--equilibrium
dynamics naturally leads to the question of whether response and
correlation obey nontrivial FD relations. In Fig.\ 3 we present the FD
plot obtained para-

\begin{figure}[t]
\begin{center}
\epsfig{file=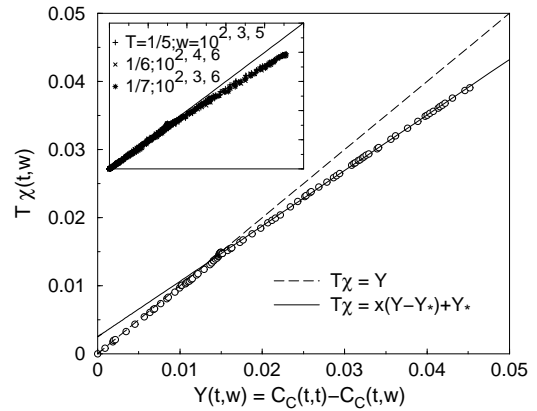, width=2.7in}
\caption
{Out--of--equilibrium FD relations in the 2D square--plaquette model.
The inset gives the FD plots for various $T$ and $w$.  The main figure
is the average of all the data.  The error bars are comparable to the
size of the symbols.  A linear fit to the second part of the curve
gives a slope $x=0.81$ and breaking point $Y_* = 0.013$.  }
\end{center}
\end{figure}

\noindent
metrically from the integrated response $\chi(t,w)$
against the difference of the connected autocorrelation $Y(t,w) \equiv
C_c(t,t)-C_c(t,w)$ for different temperatures and waiting
times.  The remarkable thing is that the FD relations for all the
values of $T$ and $w$ fall in a unique master FD plot. The FD curve
has the piecewise linear structure characteristic of discontinuous
mean--field models \cite{CuKu}, with a single breaking point
$Y_*=0.013$ and slope $x=0.81$.

Several comments are in order. The autocorrelation difference $Y(t,w)$
is playing the role of (one minus) the overlap between configurations
in this physical aging situation. The dependence of the response on
time is through this difference, $\chi(t,w)=\chi[Y(t,w)]$.  Moreover,
while $Y(t,w)$ is a non--monotonic function of $t$, it is a monotonic
function of $w$, so the FD relation can be rewritten $T \chi(Y) =
\int_0^Y X(y) dy$, where $X(y)=1$ for $y<Y_*$, and $X(y)=x$ for
$y>Y_*$. Notice also that for fixed $t$ and varying $w$, the range of
$Y$ is determined by $c(t)$, i.e., $Y(t,w) \in [0,c(t)-c^2(t)]$, and
given that for all temperatures under consideration $Y_{\rm
eq}(\infty) < Y_*$, all equilibrium curves are contained in the master
plot of Fig.\ 3.

The piecewise linear FD plot is purely a dynamical effect---the static
properties of the model are trivial. This means that violation of FDT
does not necessarily entail an RSB--like phase transition as in
disordered mean--field models \cite{Franz}. The configurations at the
plateau seem to determine the structure of the FD curve. Each of these
is composed of $c_P$ excitations, $d_P$ of which belong to an
oscillating pair. The breaking point $Y_* = d_P/2$ is given precisely
by the oscillation of all of these pairs. Higher values of $Y$
correspond to pairs of configurations mutually accessible only through
activation.  It is harder to understand the slope $x$. The fact that
it is independent of $T$ means that it is not directly related to the
entropy of configurations of isolated defects at the plateau, which
can be calculated from the hard--square model \cite{Baxter}.

The results of this work suggest that nontrivial FD relations may also
exist in the activated regime of more realistic strong glasses than
the ones studied here, and in the physical aging regime of other
systems with simple statics. While this would only be true for an
appropriately defined class of observables, consistent with the
remarks of \cite{Fielding}, for this class FD relations would be well
defined and unique. For the simple models considered here it
corresponds to observables constructed out of local energy
excitations, which are `orthogonal' to the ones responsible for the
activated relaxation, the defect concentration in the case of these
systems. It would be interesting to understand the FD ratio $X(y)$ in
terms of a geometrical, rather than equilibrium, probability
distribution for the dynamical configurations, in line with the ideas
of `Edward's measures' put forward in \cite{BarratKurchan}. And it
would also be important to extend these results to the more difficult
case of fragile systems.

We thank Josef J\"ackle and Peter Sollich for discussions. This work
was supported by EU Grant No.\ HPMF-CT-1999-00328 and the Glasstone
Fund (Oxford).

\end{multicols}

\begin{thebibliography}{}
\bibitem{Simu} Simulations were performed using continuous time Monte
Carlo\cite{Newman} for $N=10^5$--$10^6$ averaged over $20$--$50$
runs. The strength of the perturbations, $h=0.01$--$0.05$, was small
enough to be in the linear response regime.
\bibitem{Connected} FD relations are in terms of connected
correlations due to the fact that one--time quantities change with
time. This can be seen by considering a similar perturbation to the
one in the text, but restricted to $\sum_i \varepsilon_i \equiv 0$,
which leads to the same response, but only their connected correlation
functions coincide in the thermodynamic limit. Alternatively, from the
fact that FD relations for $\sigma_i$ should be the same as
those for $n_i$.
\end{thebibliography}
\end{document}